\documentclass[preprint,showpacs,preprintnumbers,amsmath,amssymb]{revtex4}
\usepackage{graphicx,epsfig}
\usepackage{amssymb}
\usepackage{amsmath}

\usepackage{graphicx}
\usepackage{dcolumn}
\usepackage{bm}
\usepackage{epsfig}
\usepackage{graphics}

\begin{document}


\title{Scattering of a Klein-Gordon particle by a
Woods-Saxon potential}

\author{Clara Rojas}
\affiliation{Centro de F\'{\i}sica IVIC Apdo 21827, Caracas 1020A,
Venezuela
}%
\author{V\'{\i}ctor M. Villalba}
\email{villalba@ivic.ve} \affiliation{Centro de F\'{\i}sica IVIC
Apdo 21827, Caracas 1020A, Venezuela
}%

\date{\today}

\begin{abstract}
We solve the Klein-Gordon equation in the presence of a spatially
one-dimensional Woods-Saxon potential. The scattering solutions are
obtained in terms of hypergeometric functions and the condition for
the existence of transmission resonances is derived. It is shown how
the zero-reflection condition depends on the shape of the potential.
\end{abstract}


\maketitle

Recently, the Woods-Saxon potential and their supersymmetric
extensions have been extensively discussed in the literature
\cite{Guo,Petrillo,Chen,Guo2,Alhaidari}. Among the advantages of
working with the Woods-Saxon potential we have to mention that, in
the one-dimensional case, the Klein-Gordon as well as the Dirac
equations are solvable in terms of special functions and therefore
the study of bound states and scattering processes becomes more
tractable. It should be mentioned that the Woods saxon potential is,
for some values of the shape parameters, a smoothed out form of the
potential barrier.

The study of low momentum scattering in the Schr\"odinger equation
in one-dimensional even potentials shows that, as momentum goes to
zero, the reflection coefficient goes to unity unless the potential
$V(x)$ supports a zero energy resonance \cite{Newton}. In this case
the transmission coefficient goes to unity, becoming a transmission
resonance \cite{Bohm}. Recently, this result has been generalized to
the Dirac equation \cite{Dombey}, showing that transmission
resonances at $k=0$ in the Dirac equation take place for a potential
barrier $V=V(x)$ when the corresponding potential well $V=-V(x)$
supports a supercritical state. The situation for short range
potentials in the Klein-Gordon equation is completely different,
here there are particle-antiparticle creation processes but no
supercritical states
\cite{Schiff,Snyder,greiner,Greiner2,Rafelski,Popov}. The absence of
supercritical states in the Klein-Gordon equation in the presence of
short-range potential interactions does not prevent the existence of
transmission resonances for given values of the potential.

The presence of transmission resonances in relativistic scalar wave
equations in external potentials has been extensively discussed in
the literature \cite{greiner,Fulling}. As a result of this
phenomenon we have that, for given values of the energy and of the
shape of the effective barrier, the probability of the transmission
coefficient reaches a maximum such as those obtained in the study of
quasinormal modes and superradiance in black hole physics.

Despite its relative simplicity, scattering processes of
relativistic scalar particles by one-dimensional potentials exhibit
the same physical properties that $s$ waves in the presence of
radial potentials, therefore the results reported in this article
can be straightforwardly extended to the radial Woods-Saxon
potential.

It is the purpose of the present article to compute the scattering
solutions of the one-dimensional Klein-Gordon equation in the
presence of a Woods-Saxon potential and show that one-dimensional
scalar wave solutions exhibit transmission resonances with a
functional dependence on the shape and strength of the potential
similar to the obtained for the Dirac equation \cite{kennedy}.

\begin{figure}[tbp]
\begin{center}
\includegraphics[width=12cm]{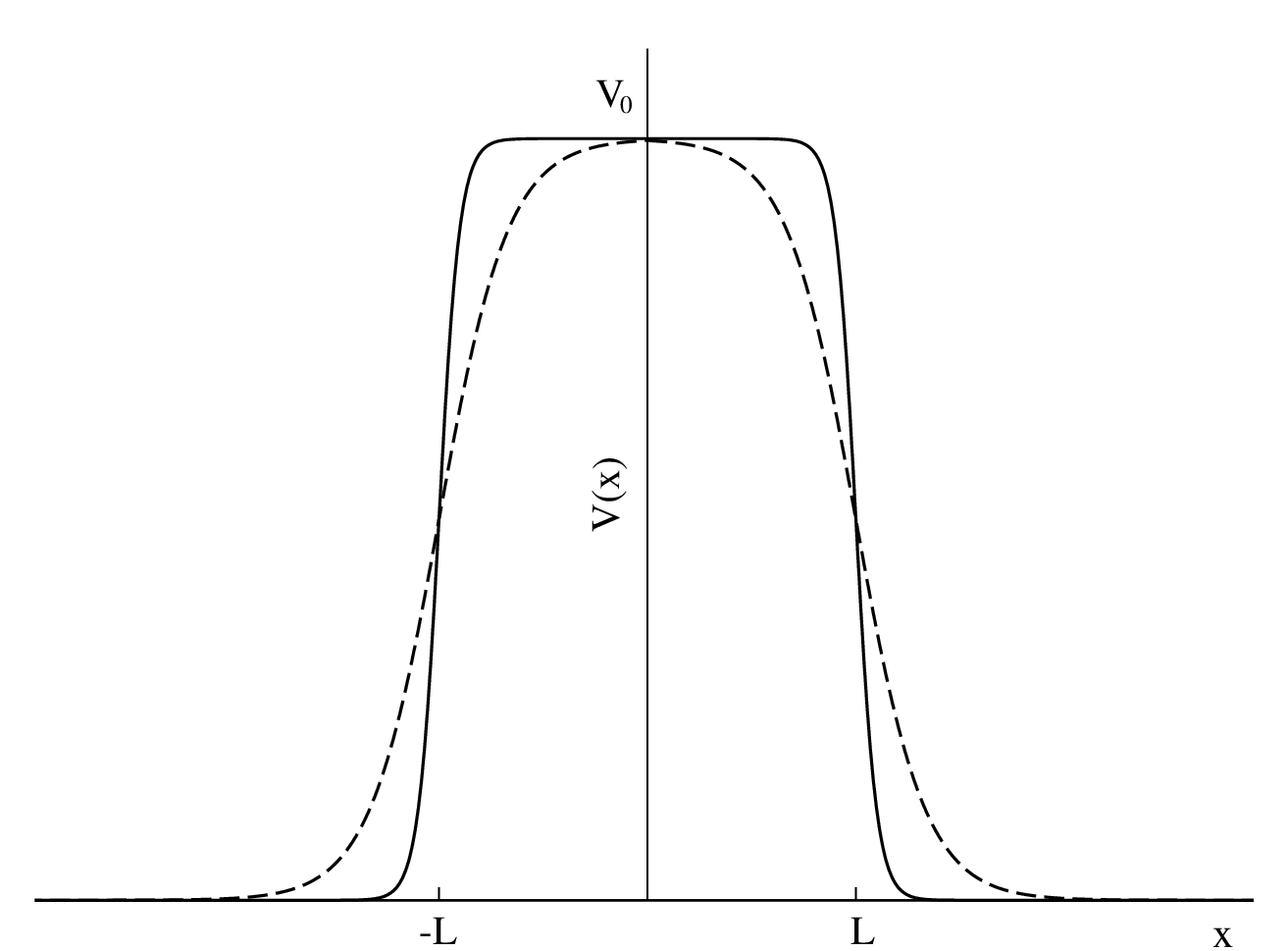}
\end{center}
\caption{{Woods-Saxon potential for $L=2$ with $a=10$ (solid line) and $a=3$
(dotted line).}}
\label{f1}
\end{figure}

The one-dimensional Klein-Gordon equation, minimally coupled to a
vector potential $A^{\mu}$ can be written as

\begin{equation}
\eta ^{\alpha \beta }(\partial _{\alpha }+ieA_{\alpha })(\partial
_{\beta }+ieA_{\beta })\phi +\phi =0 \label{ws0}
\end{equation}
where the metric $\eta ^{\alpha \beta }=diag(1,-1)$ and here and thereafter
we choose to work in natural units $\hbar =c=m=1$ \cite{greiner}

\begin{equation}
\frac{d^{2}\phi (x)}{dx^{2}}+\left[ \left( E-V(x)\right) ^{2}-1\right] \phi
(x)=0.
\end{equation}
The Woods-Saxon potential is defined as \cite{kennedy}

\begin{equation}
V(x)=V_{0}\left[ \frac{\Theta (-x)}{1+e^{-a(x+L)}}+\frac{\Theta (x)}{%
1+e^{a(x-L)}}\right] ,
\end{equation}
where $V_{0}$ is real and positive; $a>0$ and $L>0$ are also real
and positive. $\Theta (x)$ is the Heaviside step function. The form
of the Woods-Saxon potential is shown in the Fig. \ref{f1}.

From Fig. \ref{f1} one readily notices that, for a given value of the width
parameter $L$, as the shape parameter $a$ increases, the Woods-Saxon
potential reduces to a square barrier with smooth walls.

In order to consider the scattering solutions for $x<0$ with
$E^{2}>1$, we proceed to solve the differential equation

\begin{equation}
\frac{d^{2}\phi _{L}(x)}{dx^{2}}+\left[ \left( E-\frac{V_{0}}{1+e^{-a(x+L)}}%
\right) ^{2}-1\right] \phi _{L}(x)=0.  \label{ws2}
\end{equation}
On making the substitution $y=-e^{-a(x+L)}$ , Eq. (\ref{ws2}) becomes

\begin{equation}
a^{2}y\frac{d}{dy}\left[ y\frac{d\phi _{L}(y)}{dy}\right] +\left[ \left( E-%
\frac{V_{0}}{1-y}\right) ^{2}-1\right] \phi _{L}(y)=0.  \label{ws3}
\end{equation}
Putting $\phi _{L}(y)=y^{\mu }(1-y)^{-\lambda _{1}}h(y)$, Eq.
(\ref{ws3}) reduces to the hypergeometric equation

\begin{equation}  \label{ws4}
y(1-y)h^{\prime\prime}+[(1+2\mu)-(2\mu-2\lambda_1+1)y]h^{\prime}-(\mu-%
\lambda_1+\nu)(\mu-\lambda_1-\nu)h=0,
\end{equation}

\noindent where the primes denote derivatives with respect to $y$
and the parameters $\nu ,$ $k,$ $\mu ,$, $\lambda$ and $\lambda
_{1}$ are:

\begin{equation}  \label{ws5}
\nu=\frac{ik}{a}, \hspace{1cm} k=\sqrt{E^2-1,} \hspace{1cm}\mu=\frac{\sqrt{%
1-(E-V_0)^2}}{a},
\end{equation}

\begin{equation}
\lambda =\frac{\sqrt{a^{2}-4V_{0}^{2}}}{2a},\hspace{1cm}\lambda _{1}=-\frac{1%
}{2}+\lambda .  \label{ws6}
\end{equation}
The general solution of Eq. (\ref{ws4}) can be expressed in terms of
Gauss hypergeometric functions ${_{2}F_{1}}\left( \mu ,\nu, \lambda
;y\right)$ as \cite{abra}

\begin{eqnarray}
h(y) &=&D_{1}\hspace{0.05cm}{_{2}F_{1}}\left( \mu -\nu -\lambda
_{1},\mu
+\nu -\lambda _{1},1+2\mu ;y\right)  \nonumber  \label{ws7} \\
&+&D_{2}\hspace{0.05cm}y^{-2\mu }{_{2}F_{1}}\left( -\mu -\nu
-\lambda _{1},-\mu +\nu -\lambda _{1},1-2\mu ;y\right) ,
\end{eqnarray}
so

\begin{eqnarray}  \label{ws71}
\phi_L(y)&=&D_1 y^\mu(1-y)^{-\lambda_1} \hspace{0.05cm}_2F_1\left(\mu-\nu-%
\lambda_1,\mu+\nu-\lambda_1,1+2\mu;y\right)  \nonumber \\
&+&D_2 y^{-\mu}(1-y)^{-\lambda_1} \hspace{0.05cm}_2F_1\left(-\mu-\nu-%
\lambda_1,-\mu+\nu-\lambda_1,1-2\mu;y\right).
\end{eqnarray}

As $x\rightarrow -\infty $, we have that $y\rightarrow -\infty $,
and the asymptotic behavior of the solutions (\ref{ws71}) can be
determined using the asymptotic behavior of the Gauss hypergeometric
functions \cite{abra}

\begin{equation}  \label{ws8}
_2F_1(a,b,c;y)=\frac{\Gamma(c)\Gamma(b-a)}{\Gamma(b)\Gamma(c-a)}(-y)^{-a}+%
\frac{\Gamma(c)\Gamma(a-b)}{\Gamma(a)\Gamma(c-b)}(-y)^{-b}.
\end{equation}

Using Eq. (\ref{ws8}) and noting that in the limit $x\rightarrow
-\infty $, $(-y)^{\mp \nu }\rightarrow e^{\pm ik(x+L)}$ \ we have
that the asymptotic behavior of $\phi _{L}(x)$ can be written as

\begin{equation}  \label{ws9}
\phi_L(x) \rightarrow A e^{ik(x+L)}+ B e^ {-ik(x+L)},
\end{equation}

\noindent where the coefficients $A$ and $B$ in Eq. (\ref{ws9}) can be
expressed in terms of $D_{1}$ and $D_{2}$ as:

\begin{equation}  \label{ws10}
A=D_1 \frac{\Gamma(1-2\mu)\Gamma(-2\nu)(-1)^{-\mu}}{\Gamma(-\mu-\nu-%
\lambda_1)\Gamma(1-\mu-\nu+\lambda_1)}+D_2 \frac{\Gamma(1+2\mu)\Gamma(-2%
\nu)(-1)^\mu}{\Gamma(\mu-\nu-\lambda_1)\Gamma(1+\mu-\nu+\lambda_1)}.
\end{equation}

\begin{equation}  \label{ws11}
B=D_1 \frac{\Gamma(1-2\mu)\Gamma(2\nu)(-1)^{-\mu}}{\Gamma(-\mu+\nu-%
\lambda_1)\Gamma(1-\mu+\nu+\lambda_1)}+D_2 \frac{\Gamma(1+2\mu)\Gamma(2%
\nu)(-1)^\mu }{\Gamma(\mu+\nu-\lambda_1)\Gamma(1+\mu+\nu+\lambda_1)}.
\end{equation}

Now we consider the solution for $x>0$. \ In this case, the
differential equation to solve is

\begin{equation}  \label{ws12}
\frac{d^2\phi_R(x)}{dx^2}+\left[\left(E-\frac{V_0}{1+e^{a(x-L)}} \right)^2-1 %
\right]\phi_R(x)=0.
\end{equation}

The analysis of the solution can be simplified making the substitution $%
z^{-1}=1+e^{a(x-L)}$. Eq. (\ref{ws12}) can be written as

\begin{equation}  \label{ws13}
a^2 z (1-z)\frac{d}{dz}\left[z(1-z)\frac{d\phi_R(z)}{dz}\right]+\left[%
\left(E-V_0z \right)^2-1 \right]\phi_R(z)=0.
\end{equation}

Putting $\phi _{R}(z)=z^{-\nu }(1-z)^{-\mu }g(z)$, Eq. (\ref{ws13}) reduces
to the hypergeometric equation

\begin{equation}  \label{ws14}
z(1-z)g^{\prime\prime}+[(1-2\nu)-2(1-\nu-\mu)z]g^{\prime}-\left({%
\scriptstyle \frac 1 2} -\nu-\mu-\lambda\right)\left({\scriptstyle \frac 1 2}
-\nu-\mu+\lambda\right)g=0,
\end{equation}

\noindent where the primes denote derivatives with respect to $z.$ The
general solution of Eq. (\ref{ws14}) is \cite{abra}

\begin{eqnarray}
g(z) &=&d_{1}\hspace{0.05cm}_{2}F_{1}\left( {\scriptstyle\frac{1}{2}}-\nu
-\mu -\lambda ,{\scriptstyle\frac{1}{2}}-\nu -\mu +\lambda ,1-2\nu ;z\right)
\nonumber  \label{ws15} \\
&+&d_{2}z^{2\nu }\hspace{0.05cm}_{2}F_{1}\left( {\scriptstyle\frac{1}{2}}%
+\nu -\mu -\lambda ,{\scriptstyle\frac{1}{2}}+\nu -\mu +\lambda ,1+2\nu
;z\right) ,
\end{eqnarray}
so, the solution of Eq. (\ref{ws12}) can be written as:

\begin{eqnarray}
\phi _{R}(z) &=&d_{1}z^{-\nu }(1-z)^{-\mu }\hspace{0.05cm}_{2}F_{1}\left( {%
\scriptstyle\frac{1}{2}}-\nu -\mu -\lambda ,{\scriptstyle\frac{1}{2}}-\nu
-\mu +\lambda ,1-2\nu ;z\right)  \nonumber  \label{ws151} \\
&+&d_{2}z^{\nu }(1-z)^{-\mu }\hspace{0.05cm}_{2}F_{1}\left( {\scriptstyle%
\frac{1}{2}}+\nu -\mu -\lambda ,{\scriptstyle\frac{1}{2}}+\nu -\mu +\lambda
,1+2\nu ;z\right) .
\end{eqnarray}
Keeping only the solution for the transmitted wave, $d_{2}=0$ in Eq.
(\ref {ws151}), we have that in the limit $x\rightarrow \infty $,
$z$ goes to zero and  $z^{-1}\rightarrow e^{a(x-L)}$. $\phi _{R}(x)$
can be written as

\begin{equation}
\phi _{R}(x)\rightarrow d_{1}e^{ik(x-L)}  \label{ws17}
\end{equation}
The electrical current density  for the one-dimensional Klein-Gordon
equation (\ref{ws0}) is given by the expression:
\begin{equation}
\vec{j}=\frac{i}{2}\left( \phi ^{\ast }\vec{\nabla}\phi -\phi \vec{\nabla}%
\phi ^{\ast }\right)  \label{ws18}
\end{equation}
The current as $x \rightarrow -\infty$ can be decomposed as
$j_{L}=j_{in}-j_{refl}$ where $j_{in}$ is the incident current and
$j_{refl}$ is the reflected one. Analogously we have that, on the
right side, as $x \rightarrow \infty$ the current is
$j_{R}=j_{trans}$, where $j_{trans}$ is the transmitted current.

Using the reflected $j_{refl}$ and transmitted $j_{trans}$ currents,
we have that the reflection coefficient $R$, and the transmission
coefficient $T$ can be expressed in terms of the coefficients $A$,
$B$ and $d_{1}$ as:

\begin{equation}  \label{ws19}
R=\frac{j_{refl}}{j_{inc}}=\frac{|B|^2}{|A|^2},
\end{equation}

\begin{equation}  \label{ws20}
T=\frac{j_{trans}}{j_{inc}}=\frac{|d_1|^2}{|A|^2}
\end{equation}

Obviously, $R$ and $T$ are not independent, they are related via the
unitarity condition

\begin{equation}
R+T=1
\end{equation}

\begin{figure}[tbp]
{\includegraphics[width=12cm]{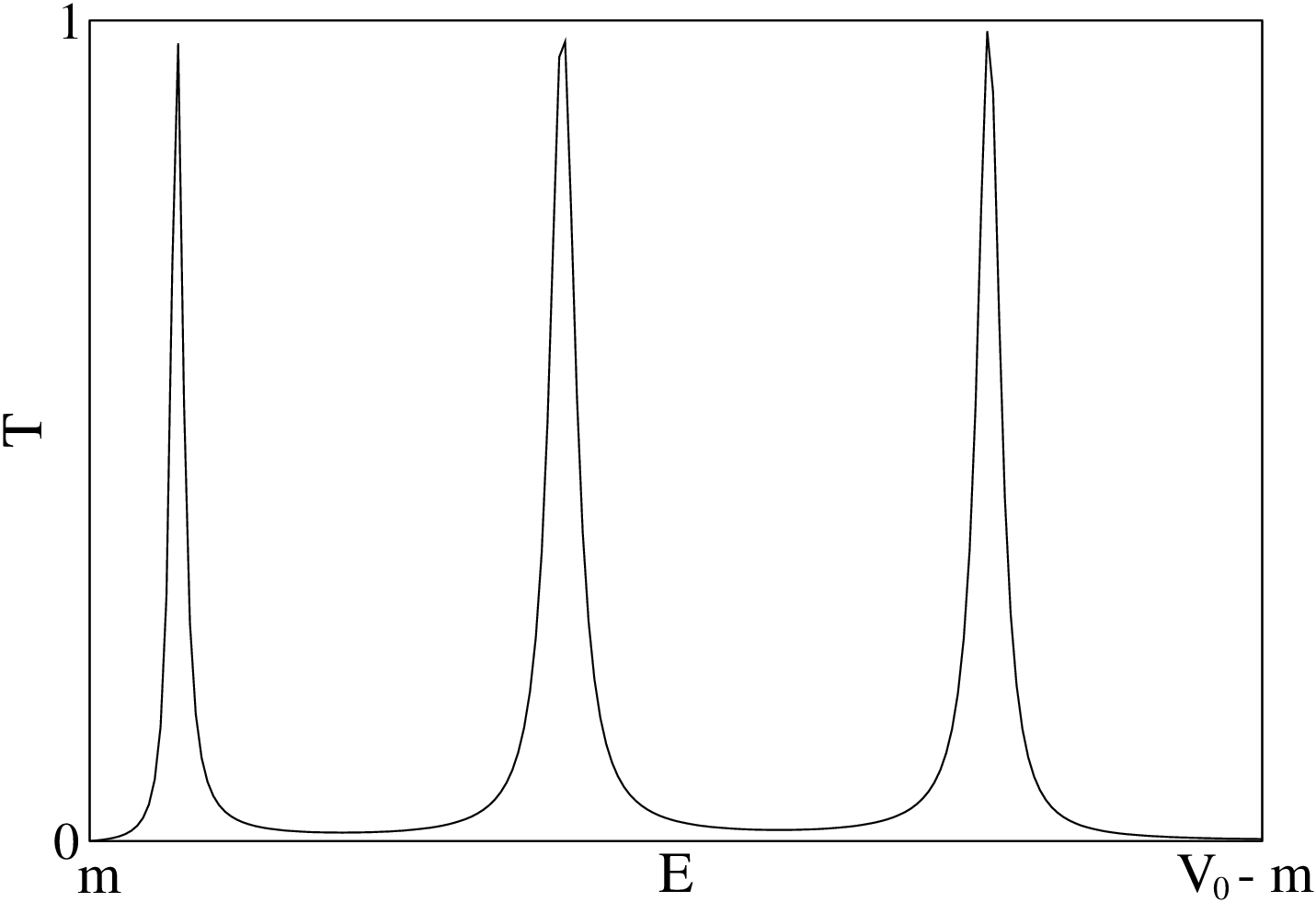}} \caption{The
transmission coefficient for the relativistic Woods-Saxon potential
barrier. The plot illustrates $T$ for varying energy, $E$ with
$a=2$, $L=2$, $m=1$ and $V_0=4$.} \label{f2}
\end{figure}

\begin{figure}[tbp]
{\includegraphics[width=12cm]{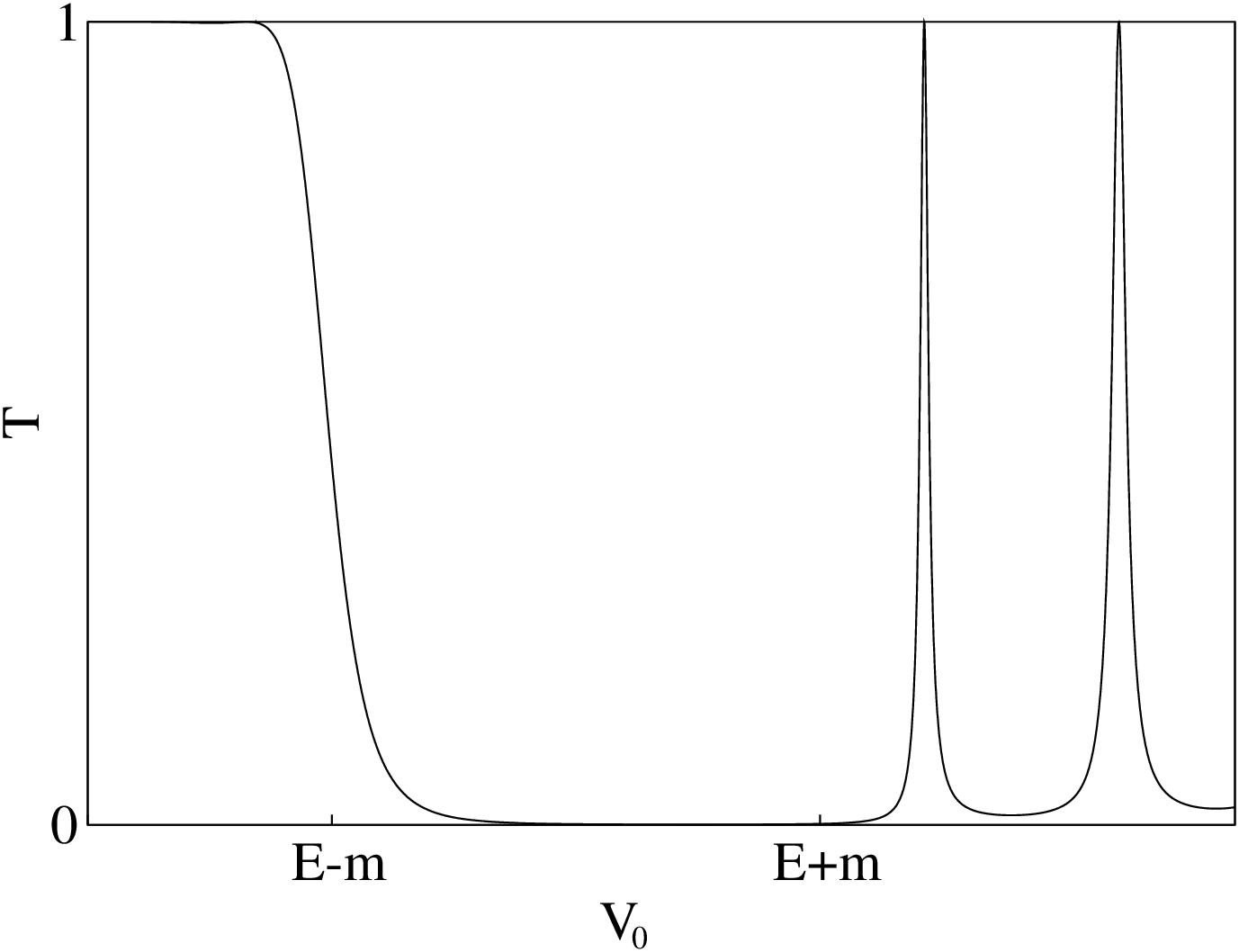}} \caption{The
transmission coefficient for the relativistic Woods-Saxon potential
barrier. The plot illustrates $T$ for varying barrier height, $V_0$
with $a=2$, $L=2$, $m=1$ and $E=2m$.} \label{f3}
\end{figure}

\begin{figure}[tbp]
{\includegraphics[width=12cm]{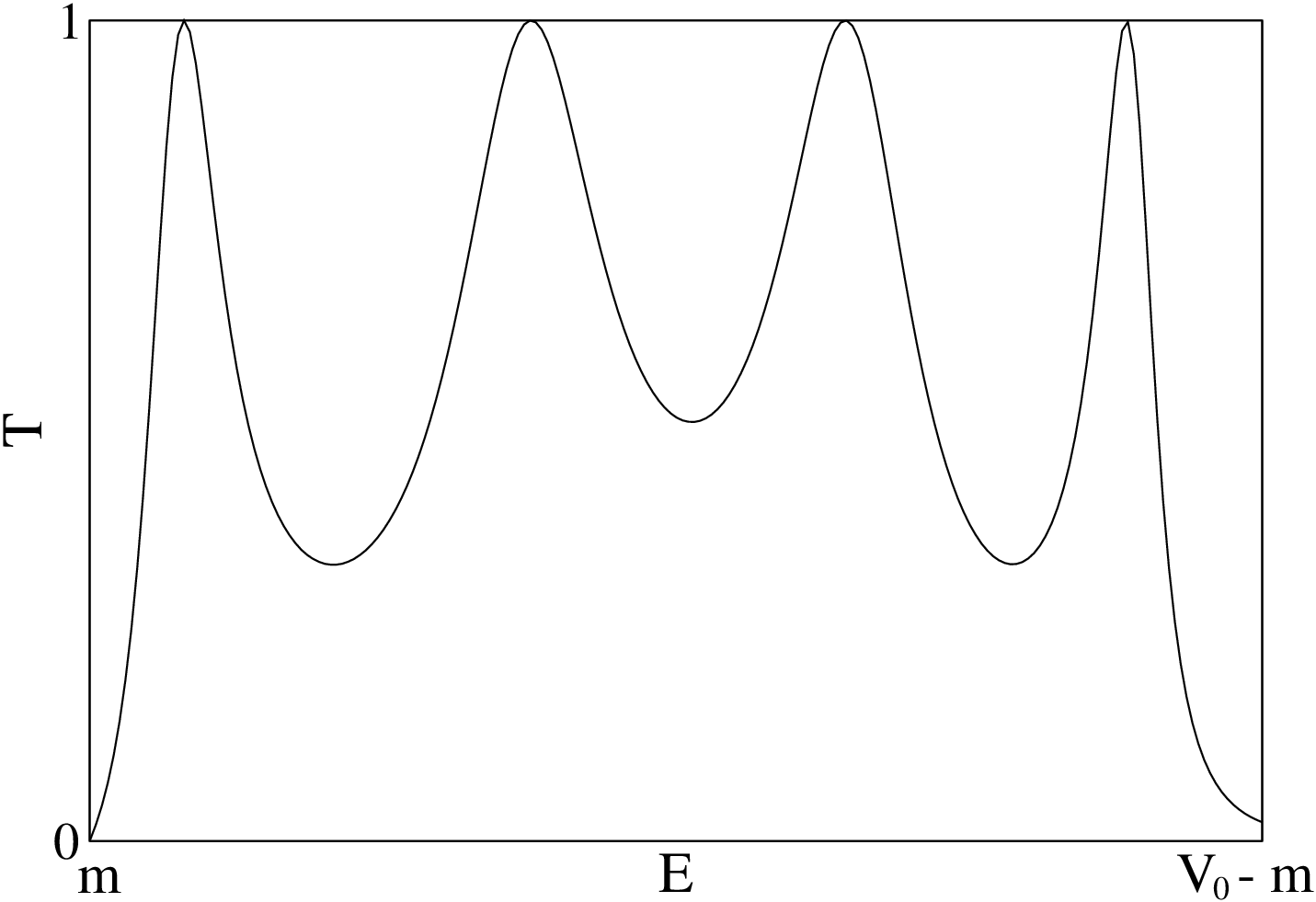}} \caption{The
transmission coefficient for the relativistic Woods-Saxon potential
barrier. The plot illustrates $T$ for varying energy, $E$ with
$a=10$, $L=2$, $m=1$ and $V_0=4$. } \label{f4}
\end{figure}

\begin{figure}[tbp]
{\includegraphics[width=12cm]{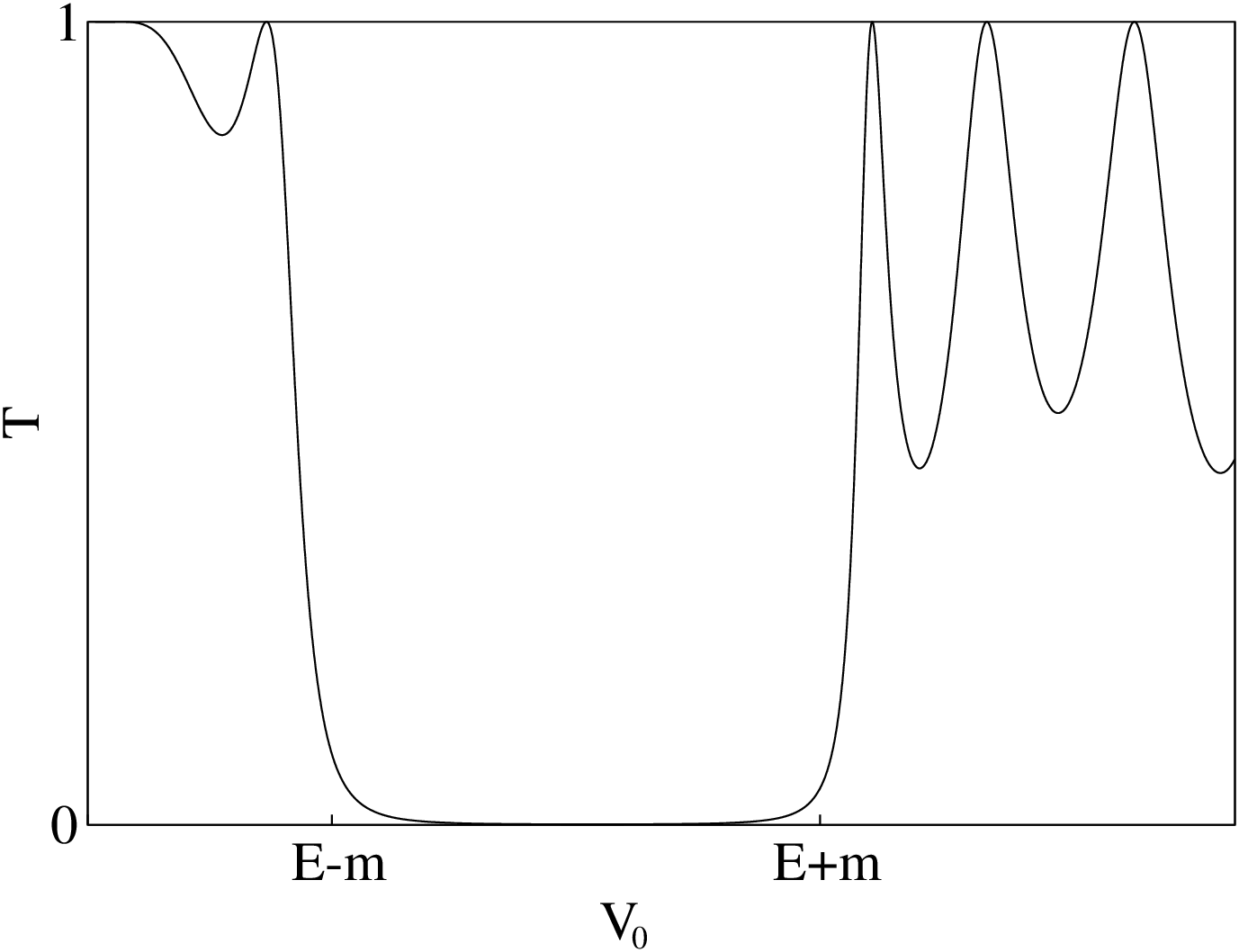}} \caption{The
transmission coefficient for the relativistic Woods-Saxon potential
barrier. The plot illustrates $T$ for varying barrier height, $V_0$
with $a=10$, $L=2$, $m=1$ and $E=2m$.} \label{f5}
\end{figure}

In order to obtain $R$ and $T$ we proceed to equate at $x=0$ the
right $\phi_{R}$ and left $\phi_{L}$ wave functions and their first
derivatives. From the matching condition we derive a system of
equations governing the dependence of coefficients $A$ and $B$ on
$d_{1}$ that can be solved numerically. Fig. \ref{f2}, \ref{f3},
show the transmission coefficients $T$, for $a=2$, $L=2$, Fig.
\ref{f4} and \ref{f5} show the transmission coefficients $T$, for
$a=10$ and $L=2$.

From Fig. \ref{f2} and Fig. \ref{f4} we can see that, analogous to
the Dirac particle, the Klein-Gordon particle exhibits transmission
resonances in the presence of the one-dimensional Woods-Saxon
potential. Fig. \ref{f2} and Fig. \ref{f4} also show that the width
of the transmission resonances depend on the shape parameter $a$
becoming wider as the Woods-Saxon potential approaches to a square
barrier.

Fig. \ref{f3} and Fig. \ref{f5} show that, as in the Dirac case, the
transmission coefficient vanishes for values of the potential
strength $E-m<V_{0}<E+m$ and transmission resonances appear for
$V_{0}>E+m$. Fig. \ref{f3} and Fig. \ref{f5} also show that the
width of the transition resonances decreases as the parameter $a$
decreases. We also conclude that, despite the fact that the behavior
of supercritical states for the Klein-Gordon equation in the
presence of short-range potentials is qualitatively different from
the one observed for Dirac particles \cite{Popov}, transmission
resonances for the one-dimensional Klein-Gordon equation possess the
same rich structure that we observe for the Dirac equation.

\acknowledgments

This work was supported by FONACIT under project G-2001000712.

\end{document}